\documentstyle[psfig]{caosp}

\begin{document}
%
\hauthor{E. Gerth {\it et al.}}
\title{Integral Representation  of the Stellar Surface Structure 
of the Magnetic Field} 
 
\author{E. Gerth
\inst{1}
\and Yu.V. Glagolevskij
\inst{2}
\and G. Scholz
\inst{1}}
\institute{Astrophysikalisches Institut Potsdam, Telegrafenberg A31, D-14473
Potsdam, Germany\\
\and Special Astrophysical Observatory of the Russian AS, Nizhnij Arkhyz
357147, Russia}
 
\date{December 11, 1997}

\maketitle

\begin{abstract}  
The surface structure of the magnetic field in stars is one part of the  
information about the surface inhomogeneities like brightness, movement or  
chemical composition, which is contained in the integral radiation flux but
distorted beyond recognition by losses of information on the topographical
arrangement over the surface because of convolution processes and partial
invisibility.
 
Therefore, the complicated processes, beginning with the representation of the
map of the surface field distribution and ending with the resulting phase-curves
of the integral magnetic field, are simulated by a computer program.  
We use this program as a tool for the analysis of the magnetic field structure
out of the really observed data in phase diagrams by variation of the parameters
and by fitting the calculated curves to the observed ones. 

\end{abstract}
 
\section{Programming of the map}  
 
{\bf Mapping} of individual features on the surface of a star out of the
integral radiation can be carried out only in the case where the linear
convolution integral transform leading to the integral flux is complete and can
be inverted. However, we have to cope with the so-called ill-posed inverse
problem, which cannot be solved correctly. The method proposed here relates
only to a limited number of parameters for the straightforward calculation, thus
offering a way to overcome this problem.
 
The map of the {\bf topographical arrangement} of the surface magnetic field
with its vectorial character is therefore constructed by matrices. The matrix
elements are defined by the spherical coordinates of the longitude and the
latitude.

The {\bf magnetic field vector} consists of three components with the unity
vectors in direction of the radius of the star (normal vector), in direction of
the longitude ($\phi$-vector), and in direction of the latitude
($\delta$-vector). A fourth component is added for a scalar  
magnitude, which can be used for different purposes (brightness, transparency,  
factor).
 
The calculation of the magnetic field components makes use of the fact that the  
linear aggregates of the potentials of {\bf point-like field sources} are
superposed linearly. Thus, the potential of a single source will be calculated
by the transform of rectangular to spherical coordinates. Then the field vector
is easily derived by the spherical gradient of the potential.
  
The advantage of the {\bf linear superposition of potentials and vector fields}
is obvious: the calculation is not limited to special source configurations.
The individual treatment of {\it monopoles} allows an arbitrary composition of
configurations up to higher multipoles. In principle, any field you like can be
represented by a row of ``monopole'' fields, the sum of charges being zero.
    
The {\bf arrangement of the sources} may be anywhere in the interior of the
stellar body. However, there could be magnetic sources placed even at the
surface, representing stellar spots like sun spots. Sources outside the star
could be positioned in companions, which influence the surface of the main star
with their fields. 
 
\section{Aspect window and convolution}
 
The {\bf aspect window} is determined by the visible hemisphere of the star. The  
program computes, for any inclination angle with respect to the rotation axis,
the projection of the elements and the limb darkening, reducing it to a
rectangular matrix of the same rank as the map.
  
The matrices of the map and of the aspect window are subjected to a {\bf matrix  
convolution}, which corresponds to the rotation of the star inclined to the line
of sight to the observer. There are a series of geometrical aspects with a
number of steps, determined by the rank of the matrix. In the present version
of the program the highest rank is 104.
 
The {\bf convolution algorithm} is the core of the program. It is multivalent
and can be used also for other surface quantities.
 
\section{Physical problems of the integral magnetic field} 
 
Usually we interpret the effective stellar magnetic field by the {\bf Zeeman 
displacement} of the gravity centers of the line profiles of oppositely
polarized light. Since the light comes from the visible hemisphere of the star,
all parts of the surface contribute differently to the whole profile and its
resulting spectral position.
  
The {\bf effective magnetic field} $B_{\rm eff}$ is not simply a mean value but
already the result of weighting and convolution of the radiation flux
containing the magnetic field information in the form and position of the
profiles of all surface elements. 

We relate to the fact, that the {\bf center of gravity} of the profiles is
given by the mean of the centres weighted by the profile integrals. Thus, we
weigh the magnetic field vector of all surface elements with their spherical
projection onto the line of sight including the limb darkening and integrate
them over the visible hemisphere. The magnetic field is treated by the same
procedure as the brightness, regarding only the vectorial
character of the magnetic field.
  
The magnetic field is also influenced by the inhomogeneous distribution of
{\bf chemical elements} over the star's surface. This acts like a transmission
factor for every surface element and undergoes likewise the weighting of
profiles. The fourth component of a scalar magnitude takes the map of the
chemical distribution into account.   
 
\section{Graphical representation and storage of maps and results} 
 
The usefulness of the graphical representation of the calculation is beyond any  
discussion. In the case of the present program, computer graphics have already  
served at the first stages of development for controlling and testing purposes.
 
The {\bf map} of the magnetic field structure is represented by colored
isolines. 

The {\bf rotation curves} are computed and drawn on to the map, so that it is
easy to see how the topographical features act onto the curves.  

The calculated maps and phase {\bf curves are stored} as files on disk and may
be recalled repeatedly for different calculations and graphical representations
in connection with the phase-arranged observational data of real stellar
objects, which are plotted in the phase diagrams corresponding to the maps of
the magnetic field and/or the chemical distribution.
 
The program coordinates the magnetic field structure to the {\bf chemical map}
(derived by other authors and methods) but does not analyze its distribution
structure.   
 
\section{The use of the program} 
 
The program is used for the investigation of the characteristic {\bf shapes of
curves}, as have been observed in magnetic stars. Examples can be seen in the
papers about the stars CU Vir and $\alpha^{2}$~CVn.

A {\bf catalogue} of maps and resulting curves with systematic variation of
parameters will be included in a data bank for comparison with observed curves.
 
The program will be applied to real early type magnetic stars in order to 
determine the topographical positions of the poles and to discover their
magnetic structure as the basis for the explanation of the connection between
{\bf stellar magnetism and chemical composition}. 

\end{document}